\documentclass{emulateapj}

\begin{document}
\title{Reply to Comment on ``Spectra of strong magnetohydrodynamic turbulence from high-resolution simulations''}
\author{A. Beresnyak}
\affiliation{Nordita, KTH Royal Institute of Technology and Stockholm University, SE-10691 Stockholm, Sweden}

\begin{abstract}
In a Comment by \citet{Perez2014} it is claimed that recently published simulations of \citet{B14a} are grossly
underresolved, compared to theirs, and that \citet{B14a} failed to estimate numerical error. Both claims 
are contrary to the fact. Firstly, as far as numerical resolution is concerned,
Beresnyak (2014) was using $k_{\rm max}\eta_{41}$=1.06 resolution criterion, while Perez et al has been using $k_{\rm max}\eta_{41}$=0.8.
Obviously, $1.06>0.8$. Secondly, \cite{B14a} have estimated numerical
error and found it to be below $3\times 10^{-3}$, which is properly explained in the paper. On close inspection
of the Comment I have not found a single numerical value or parameter pertaining to the criticized paper \citep{B14a},
and it is completely unclear how the Authors came to their conclusions.
\end{abstract}

\maketitle

Numerical simulations of MHD turbulence in strong mean field with pseudospectral code date back to early 2000s \citep{CV00}.
Recently there has been a debate regarding the spectral slope in high-resolution simulations between J.Perez et al and A. Beresnyak.
Perez et al has been repeatedly 
bringing numerical inaccuracies as the main source of disagreement \citep[see, e.g., ][]{Perez2012,Perez2014}.
The latest Comment \citep{Perez2014} describe Beresnyak's simulations
as ``drastically unresolved'' and their own simulations as numerically accurate. 
This is quite surprising, considering that Beresnyak's simulations are, in fact, better resolved than those reported by Perez et al. 

While carefully reading the Comment I have found no parameters from the criticized paper, \cite{B14a}. 
The Author's claim seems to be completely
arbitrary. They are trying to allege that some of their own grossly underresolved simulations presented on their Figs.~1 and 2
have anything to do with \cite{B14a}. This is simply not the case.

The main resolution criterion in turbulence simulations is based on a ratio of Kolmogorov (dissipation) scale $\eta$ to the grid
scale. In pseudospectral simulations of both groups the box size is $2\pi$ and the grid size is $2\pi/N$, where $N$ is the mesh size.
Another useful quantity is a maximum wavenumber $k_{\rm max}$, equal to $N/3$ in the $2/3$ dealiased
simulations of both groups. It follows that $k_{\rm max}=2\pi/(3\Delta)$ and both the ratio $\eta/\Delta$ and the product of 
$k_{\rm max}\eta$ can be used as a numerical resolution parameter, with higher parameter corresponding to higher numerical
accuracy.

The Kolmogorov scale $\eta$ is itself a function of the model. This, however, is not a problem, as long as
the same definition is used for comparison. \cite{Perez2012} designate $\eta_{41}=(\nu^3/\epsilon)^{1/4}$, which is the classic
Kolmogorov scale and their Fig.~9 indicate that $k_{\rm max}\eta_{41}=0.8$ in their case. This can be independently verified by using
parameter $\nu=1/{\rm Re}$ from Table I, simulations RB1a, RB2a and RB3a
and parameter $\epsilon=0.15$ from page 8. \cite{B14a}, however,
uses $k_{\rm max}\eta_{41}=1.06$, which is a better resolution that corresponds to higher numerical accuracy. How the Authors
of the Comment concluded that they have better resolution is totally puzzling. They do not mention the resolution of \cite{B14a}
for that matter.

The slightly reduced numerical accuracy of \cite{Perez2012} is not the biggest problem of their paper, however.
As we noted earlier in \cite{B13comm}, this paper have severe methodological flaw of claiming correspondance
between theory and strange numerical ``measurement''.    
On Fig.~8 of \cite{Perez2012} it is claimed that the measured length of the inertial range follows
 scaling from \cite{boldyrev2006}, namely $Re^{2/3}$. On close inspection, however, it is evident
that the Authors calculated ``datapoints'' by the formula  $0.025 \epsilon^{2/9} \Lambda^{-1/9} \nu^{-2/3}$,
where $Re=1/\nu$ and $0.025$ is a number, arbitrarily chosen by the Authors. 
Using $\epsilon$ and $\Lambda$ quoted in \cite{Perez2012} the product $\epsilon^{2/9} \Lambda^{-1/9}$
 can be approximated as $0.517$, after which the dependence $0.0129 {\rm Re}^{2/3}$ reproduces the Author's plot on the bottom of Fig.~8. 
Some time have passed after publication of \cite{B13comm}, but the claim from \cite{Perez2012}
 have not been recalled by the Authors yet, which is deeply troubling, in my opinion.

In the end of their Comment, the Authors claimed that \cite{B14a} have failed to perform numerical convergence study and estimate
numerical error. This is contrary to the fact, see \cite{B14a}, page 2.

\def\apj{{\rm ApJ}}           
\def\apjl{{\rm ApJ }}          
\def\apjs{{\rm ApJ }}          
\def\grl{{\rm GRL }}
\def\aap{{\rm A\&A } }
\def\mnras{{\rm MNRAS } }
\def\physrep{{\rm Phys. Rep. } }               
\def\prl{{\rm Phys. Rev. Lett.}} 
\def\pre{{\rm Phys. Rev. E}} 
\bibliography{all}

\begin{thebibliography}{6}%
\makeatletter
\providecommand \@ifxundefined [1]{%
 \@ifx{#1\undefined}
}%
\providecommand \@ifnum [1]{%
 \ifnum #1\expandafter \@firstoftwo
 \else \expandafter \@secondoftwo
 \fi
}%
\providecommand \@ifx [1]{%
 \ifx #1\expandafter \@firstoftwo
 \else \expandafter \@secondoftwo
 \fi
}%
\providecommand \natexlab [1]{#1}%
\providecommand \enquote  [1]{``#1''}%
\providecommand \bibnamefont  [1]{#1}%
\providecommand \bibfnamefont [1]{#1}%
\providecommand \citenamefont [1]{#1}%
\providecommand \href@noop [0]{\@secondoftwo}%
\providecommand \href [0]{\begingroup \@sanitize@url \@href}%
\providecommand \@href[1]{\@@startlink{#1}\@@href}%
\providecommand \@@href[1]{\endgroup#1\@@endlink}%
\providecommand \@sanitize@url [0]{\catcode `\\12\catcode `\$12\catcode
  `\&12\catcode `\#12\catcode `\^12\catcode `\_12\catcode `\%12\relax}%
\providecommand \@@startlink[1]{}%
\providecommand \@@endlink[0]{}%
\providecommand \url  [0]{\begingroup\@sanitize@url \@url }%
\providecommand \@url [1]{\endgroup\@href {#1}{\urlprefix }}%
\providecommand \urlprefix  [0]{URL }%
\providecommand \Eprint [0]{\href }%
\providecommand \doibase [0]{http://dx.doi.org/}%
\providecommand \selectlanguage [0]{\@gobble}%
\providecommand \bibinfo  [0]{\@secondoftwo}%
\providecommand \bibfield  [0]{\@secondoftwo}%
\providecommand \translation [1]{[#1]}%
\providecommand \BibitemOpen [0]{}%
\providecommand \bibitemStop [0]{}%
\providecommand \bibitemNoStop [0]{.\EOS\space}%
\providecommand \EOS [0]{\spacefactor3000\relax}%
\providecommand \BibitemShut  [1]{\csname bibitem#1\endcsname}%
\let\auto@bib@innerbib\@empty
\bibitem [{\citenamefont {{Beresnyak}}(2014)}]{B14a}%
  \BibitemOpen
  \bibfield  {author} {\bibinfo {author} {\bibfnamefont {A.}~\bibnamefont
  {{Beresnyak}}},\ }\href {\doibase 10.1088/2041-8205/784/2/L20} {\bibfield
  {journal} {\bibinfo  {journal} {\apjl}\ }\textbf {\bibinfo {volume} {784}},\
  \bibinfo {eid} {L20} (\bibinfo {year} {2014})}\BibitemShut {NoStop}%
\bibitem [{\citenamefont {{Cho}}\ and\ \citenamefont
  {{Vishniac}}(2000)}]{CV00}%
  \BibitemOpen
  \bibfield  {author} {\bibinfo {author} {\bibfnamefont {J.}~\bibnamefont
  {{Cho}}}\ and\ \bibinfo {author} {\bibfnamefont {E.~T.}\ \bibnamefont
  {{Vishniac}}},\ }\href {\doibase 10.1086/309127} {\bibfield  {journal}
  {\bibinfo  {journal} {\apj}\ }\textbf {\bibinfo {volume} {538}},\ \bibinfo
  {pages} {217} (\bibinfo {year} {2000})},\ \Eprint
  {http://arxiv.org/abs/arXiv:astro-ph/0003404} {arXiv:astro-ph/0003404}
  \BibitemShut {NoStop}%
\bibitem [{\citenamefont {{Perez}}\ \emph {et~al.}(2012)\citenamefont
  {{Perez}}, \citenamefont {{Mason}}, \citenamefont {{Boldyrev}},\ and\
  \citenamefont {{Cattaneo}}}]{Perez2012}%
  \BibitemOpen
  \bibfield  {author} {\bibinfo {author} {\bibfnamefont {J.~C.}\ \bibnamefont
  {{Perez}}}, \bibinfo {author} {\bibfnamefont {J.}~\bibnamefont {{Mason}}},
  \bibinfo {author} {\bibfnamefont {S.}~\bibnamefont {{Boldyrev}}}, \ and\
  \bibinfo {author} {\bibfnamefont {F.}~\bibnamefont {{Cattaneo}}},\ }\href
  {\doibase 10.1103/PhysRevX.2.041005} {\bibfield  {journal} {\bibinfo
  {journal} {Physical Review X}\ }\textbf {\bibinfo {volume} {2}},\ \bibinfo
  {eid} {041005} (\bibinfo {year} {2012})},\ \Eprint
  {http://arxiv.org/abs/1209.2011} {arXiv:1209.2011 [astro-ph.SR]} \BibitemShut
  {NoStop}%
\bibitem [{\citenamefont {{Perez}}\ \emph {et~al.}(2014)\citenamefont
  {{Perez}}, \citenamefont {{Mason}}, \citenamefont {{Boldyrev}},\ and\
  \citenamefont {{Cattaneo}}}]{Perez2014}%
  \BibitemOpen
  \bibfield  {author} {\bibinfo {author} {\bibfnamefont {J.~C.}\ \bibnamefont
  {{Perez}}}, \bibinfo {author} {\bibfnamefont {J.}~\bibnamefont {{Mason}}},
  \bibinfo {author} {\bibfnamefont {S.}~\bibnamefont {{Boldyrev}}}, \ and\
  \bibinfo {author} {\bibfnamefont {F.}~\bibnamefont {{Cattaneo}}},\
  }\href@noop {} {\bibfield  {journal} {\bibinfo  {journal} {ArXiv e-prints}\ }
  (\bibinfo {year} {2014})},\ \Eprint {http://arxiv.org/abs/1409.8106}
  {arXiv:1409.8106 [astro-ph.SR]} \BibitemShut {NoStop}%
\bibitem [{\citenamefont {{Beresnyak}}(2013)}]{B13comm}%
  \BibitemOpen
  \bibfield  {author} {\bibinfo {author} {\bibfnamefont {A.}~\bibnamefont
  {{Beresnyak}}},\ }\href@noop {} {\bibfield  {journal} {\bibinfo  {journal}
  {ArXiv e-prints}\ } (\bibinfo {year} {2013})},\ \Eprint
  {http://arxiv.org/abs/1301.7425} {arXiv:1301.7425 [astro-ph.GA]} \BibitemShut
  {NoStop}%
\bibitem [{\citenamefont {{Boldyrev}}(2006)}]{boldyrev2006}%
  \BibitemOpen
  \bibfield  {author} {\bibinfo {author} {\bibfnamefont {S.}~\bibnamefont
  {{Boldyrev}}},\ }\href {\doibase 10.1103/PhysRevLett.96.115002} {\bibfield
  {journal} {\bibinfo  {journal} {\prl}\ }\textbf {\bibinfo {volume} {96}},\
  \bibinfo {pages} {115002} (\bibinfo {year} {2006})}\BibitemShut {NoStop}%
\end{thebibliography}%

\end{document}